\documentclass[
	aps,
	prd,
	reprint,
	amssymb,amsmath,
	nofootinbib,
	superscriptaddress,
	notitlepage,
]{revtex4-1} %% prd

\usepackage{graphicx}% Include figure files
\usepackage{dcolumn}% Align table columns on decimal point
\usepackage{bm}% bold math
\usepackage{color}
\usepackage{xspace}
\usepackage{ulem}
\newcommand{\fermi}{\emph{Fermi}\xspace}
\newcommand{\planck}{\emph{Planck}\xspace}

\newcommand{\lat}{\emph{Fermi}-LAT\xspace}

\newcommand{\gr}{$\gamma$-ray\xspace}
\newcommand{\grs}{$\gamma$ rays\xspace}
\newcommand{\ionnew}[2]{\rm{#1\,\textsc{\lowercase{#2}}}}
\newcommand{\hi}{\ionnew{H}{I}\xspace}
\newcommand{\hii}{\ionnew{H}{II}\xspace}
\newcommand{\htwo}{$\rm H_2$\xspace}

\newcommand{\kms}{\ensuremath{\rm km\;s^{-1}}}
\renewcommand{\deg}{\ensuremath{^{\circ}}\xspace}

\newcommand\aap{Astron. Astrophys.}
\newcommand\aj{Astron. J.}
\newcommand\apjs{Astrophys. J. Suppl. Ser.}
\newcommand\araa{Ann. Rev. Astron. Astrophys.}
\newcommand\mnras{Mon. Not. R. Astron. Soc.}

\newcommand\ssr{Space Sci. Rev.}
\newcommand\pasj{Publ. Astron. Soc. Jpn.}
\newcommand\aaps{Astron. Astrophys. Suppl. Ser.}
\newcommand\pasa{Publ. Astron. Soc. Aust.}

\begin{document}
\title{Probing local cosmic rays using \emph{Fermi}-LAT observation of a mid-latitude region in the third Galactic quadrant}

\author{Zhao-Qiang~Shen}
\affiliation{Key Laboratory of Dark Matter and Space Astronomy, Purple Mountain Observatory, Chinese Academy of Sciences, Nanjing 210008, China}
\affiliation{University of Chinese Academy of Sciences, Beijing 100012, China}
\author{Xiaoyuan~Huang}
\affiliation{Key Laboratory of Dark Matter and Space Astronomy, Purple Mountain Observatory, Chinese Academy of Sciences, Nanjing 210008, China}
\author{Qiang~Yuan}
\affiliation{Key Laboratory of Dark Matter and Space Astronomy, Purple Mountain Observatory, Chinese Academy of Sciences, Nanjing 210008, China}
\affiliation{School of Astronomy and Space Science, University of Science and Technology of China, Hefei, Anhui 230026, China}
\author{Yi-Zhong~Fan}
\email[]{yzfan@pmo.ac.cn}
\affiliation{Key Laboratory of Dark Matter and Space Astronomy, Purple Mountain Observatory, Chinese Academy of Sciences, Nanjing 210008, China}
\affiliation{School of Astronomy and Space Science, University of Science and Technology of China, Hefei, Anhui 230026, China}
\author{Da-Ming~Wei}
\email[]{dmwei@pmo.ac.cn}
\affiliation{Key Laboratory of Dark Matter and Space Astronomy, Purple Mountain Observatory, Chinese Academy of Sciences, Nanjing 210008, China}
\affiliation{School of Astronomy and Space Science, University of Science and Technology of China, Hefei, Anhui 230026, China}

\date{\today}
%%%%%%%%%%%%%%%%%%%%%%%%%%%%% ABSTRACT %%%%%%%%%%%%%%%%%%%%%%%%%%%%%%%
\begin{abstract}
The $\gamma$-ray observation of interstellar gas provides a unique way to probe the cosmic rays (CRs) outside the solar system.
In this work, we use an updated version of \emph{Fermi}-LAT data and recent multi-wavelength tracers of interstellar gas to re-analyze a mid-latitude region in the third Galactic quadrant and estimate the local CR proton spectrum.
Two $\gamma$-ray production cross section models for $pp$ interaction, the commonly used one from Kamae et~al. (2006) and the up-to-date one from Kafexhiu et~al. (2014), are adopted separately in the analysis.
Both of them can well fit the emissivity and the derived proton spectra roughly resemble the direct measurements from AMS-02 and \emph{Voyager 1}, but rather different spectral parameters are indicated.
A break at $4\pm1~{\rm GeV}\;c^{-1}$ is shown if the cross section model by Kamae et~al. (2006) is adopted.
The resulting spectrum is $\lesssim 20\%$ larger than the AMS-02 observation above 15~GeV and consistent with the de-modulated spectrum within $2\%$.
The proton spectrum based on the cross section model of Kafexhiu et~al. (2014) is about $1.4-1.8$ times that of AMS-02 at $2-100~\rm GeV$, however the difference decreases to 20\% below 10~GeV with respect to the de-modulated spectrum.
A spectral break at $20\pm11~{\rm GeV}\;c^{-1}$ is required in this model.
An extrapolation down to 300~MeV is performed to compare with the observation of \emph{Voyager 1}, and we find a deviation of $\lesssim 2.5\sigma$ for both the models.
In general, an approximately consistent CR spectrum can be obtained using \gr observation nowadays, but we still need a better $\gamma$-ray production cross section model to derive the parameters accurately.
\end{abstract}

\maketitle

%%%%%%%%%%%%%%%%%%%%%%%%%%%%% CONTENTS %%%%%%%%%%%%%%%%%%%%%%%%%%%%%%%
\section{Introduction}\label{sec:introduction}
Cosmic rays (CRs) play a vital role in the Galactic ecosystem, because they heat and ionize the interstellar gas, and provide an additional support against the gravitational force together with the magnetic field~\citep{Ferriere2001_VK2AK,Grenier2015_TZ3H7}.
Nowadays, there are some experiments aiming at collecting the CR particles, however due to the solar modulation, the intrinsic CR spectra in local interstellar space (LIS) below $\sim 10$~GeV/nuc can not be measured directly near the Earth~\citep{Gleeson1968_Q5UQ4}.
The \emph{Voyager 1} and \emph{Voyager 2} crossed the heliopause on 2012 August 25~\citep{Stone2013_G966N} and 2018 November 5 respectively, and started to measure the CR spectra outside the heliosphere~\citep{Stone2013_G966N,Cummings2016_W98KL}, which are thought to be the same as the LIS ones~\citep{Kota2014_Y6QEY}.
But the LIS proton spectrum from $0.35~\rm GeV$ to $\sim 10~\rm GeV$ is still not available right now~\citep{Cummings2016_W98KL}.

The interaction of CRs with the interstellar gas will produce \gr photons.
On one hand, these \grs can be a useful tracer of total gas column density~\citep{Lebrun1982_Q69MG,Grenier2005_636PV,Ade2015_UUM64,Mizuno2016_T9C4F,Abrahams2017_Y8A8J,Remy2017_7VYY4,Remy2018_VVFJ9,Remy2018_2ZF33}, since the \grs are transparent to the interstellar medium (ISM) and also independent of the chemical and thermodynamic state.
On the other hand, \gr observation provides a unique way to probe the Galactic CRs outside the solar system.
Particularly, the observation of distant gas reflects the CR spectra there, which will shed light on the origin and propagation of CR or even help to find the site of CR acceleration~\citep{Aharonian2001_86EBV,Casanova2010_Z5YFT}.

The \emph{Fermi Gamma-ray Space Telescope} (\fermi) is launched on 2008 June 11, with a pair-conversion telescope, Large Area Telescope (LAT), on board~\citep{FermiLAT2009}.
Thanks to its unprecedented sensitivity and accurate calibration~\citep{Abdo2009_PWKQ8,Abdo2009_FNCQF,Ackermann2012_CKWEC}, a plenty of researches have been done to constrain the CR spectra elsewhere in the Galaxy~\citep{Abdo2009_3VMFM,Abdo2010_ALWUR,Ackermann2011_VUC80,Neronov2012_U4LP7,Ackermann2012_BRHCL,Ackermann2012_KKTUP,Ackermann2012_LN81J,Yang2014_CNNFV,Abrahams2015_FNCQF,Yang2015_GQ8M5,Tibaldo2015_BGTRD,Casandjian2015_K3EGW,Yang2016_DKYML,FermiGDE2016,Neronov2017_5UWT9,Shen2018_L5GEE,Aharonian2018_PRU3T}.
Interstellar gas in the mid-Galactic latitude region is a favorable target to study the LIS CRs, because the gas there is mostly not far from the sun~\citep{Abdo2009_3VMFM}.
The first \gr analysis of local \hi gas in \fermi era is performed in~\cite{Abdo2009_3VMFM} and it is found to be consistent with the {\tt Galprop} prediction.
Further efforts aim at deriving the LIS CR spectra using the \gr observation of all mid-Galactic regions down to 60~MeV~\citep{Casandjian2015_K3EGW,Strong2015_NELQ4}.
The results are quite close to the PAMELA spectrum after the solar modulation correction, when the systematic uncertainties are considered.
Giant molecular clouds (GMCs) in the Gould Belt are also adopted to probe the LIS CR, because these clouds are nearby and bright in \gr sky.
Some nearby GMCs have been analyzed~\cite{Neronov2012_U4LP7,Yang2014_CNNFV,Neronov2017_5UWT9} and their emissivities are found to be similar, suggesting the \gr emission is mainly from the passive interaction with the Galactic CR sea which is also confirmed in~\citep{Ackermann2012_KKTUP,Peng2019_VDRA3}.
The CR spectrum can therefore be obtained with the emissivities of \htwo, however point source contamination might be a problem in the low energy range due to their relatively small size~\citep{Yang2014_CNNFV}.

Over the last few years, the quality of \lat data has been improved, which not only provides a larger effective area particularly in the lower energy range, but also reduces the instrumental systematic uncertainties~\citep{Atwood2013_C82D0}.\footnote{\url{https://fermi.gsfc.nasa.gov/ssc/data/analysis/LAT_caveats.html}}
New multi-wavelength observations of ISM are available, e.g. the \hi survey from~\cite{HI4PI2016}, the dust opacity and extinction from~\citep{PlanckOpacity2016,PlanckAvq2016}.
Furthermore, the \gr production cross section model for $pp$ interaction is updated in~\cite{Kafexhiu2014_R6K7Z}.
Taking advantage of the updated observations and tools, we revisit the analysis of a mid-Galactic latitude region in the third quadrant which has be done in~\cite{Abdo2009_3VMFM}.
We choose this region because local atomic hydrogen dominates the gas column density in it~\citep{Abdo2009_FNCQF}, which enables us to directly calculate the number of atoms along the line of sight and therefore is less prone to the uncertainty of the dark gas and CO-to-\htwo conversion factor.
Comparing to the previous work in~\citep{Casandjian2015_K3EGW}, we perform our analysis in a relatively clean region, use the updated ISM tracers to estimate the gas column density and more complete \lat 8-year source catalog to reduce the point source contamination.

In this paper, data reduction, including the template generation and the analysis procedure, is described in Sec.~\ref{sec:analysis}.
In the Sec.~\ref{sec:result_discussion}, the \gr spectrum and its systematic uncertainty are presented.
We then extract the LIS proton spectrum using either the latest cross section model from \cite{Kafexhiu2014_R6K7Z} or the popular one from \cite{Kamae2006_UZAJA}, and compare them with the direct measurements of AMS-02~\citep{Aguilar2015_DRNZF} and \emph{Voyager 1}~\citep{Cummings2016_W98KL}.
Finally, a summary is given in Sec.~\ref{sec:summary}.

\section{Data analysis}\label{sec:analysis}
\subsection{\gr data}\label{sec:analysis:gr_data}
The \lat P8R3 data, based on the most recent iteration of the event-level analysis, are released recently.
In this data version, the leak of charged particles through the scintillating ribbons is removed, and therefore the anisotropy problem of the background model in the previous version is solved~\citep{Bruel2018_1LQ9V}.
We choose the Clean event class of P8R3 data.\footnote{\url{ftp://legacy.gsfc.nasa.gov/fermi/data/lat/weekly/photon/}}
By using this data set, we can suppress the residual CR background at a reasonable cost of data.
Photons observed from 2008 August 4 to 2018 November 22 (\fermi Mission Elapsed Time (MET) from 239557417 to 564539821) with energy between 75~MeV and 100~GeV are selected.
We further exclude the reconstructed zenith angles over 85\deg to reduce the contamination from the Earth's limb, and then apply the recommended quality-filter cut $\rm (DATA\_QUAL>0)\&\&(LAT\_CONFIG==1)$, which removes the events collected outside science mode or during the time interval when either a solar flare or particle event happens.
The events between July 14 and September 13 in each year are also excluded in order to remove the emission from the Sun in the region of interest (ROI) defined below.

\begin{figure*}
    \centering
    \includegraphics[width=0.48\textwidth]{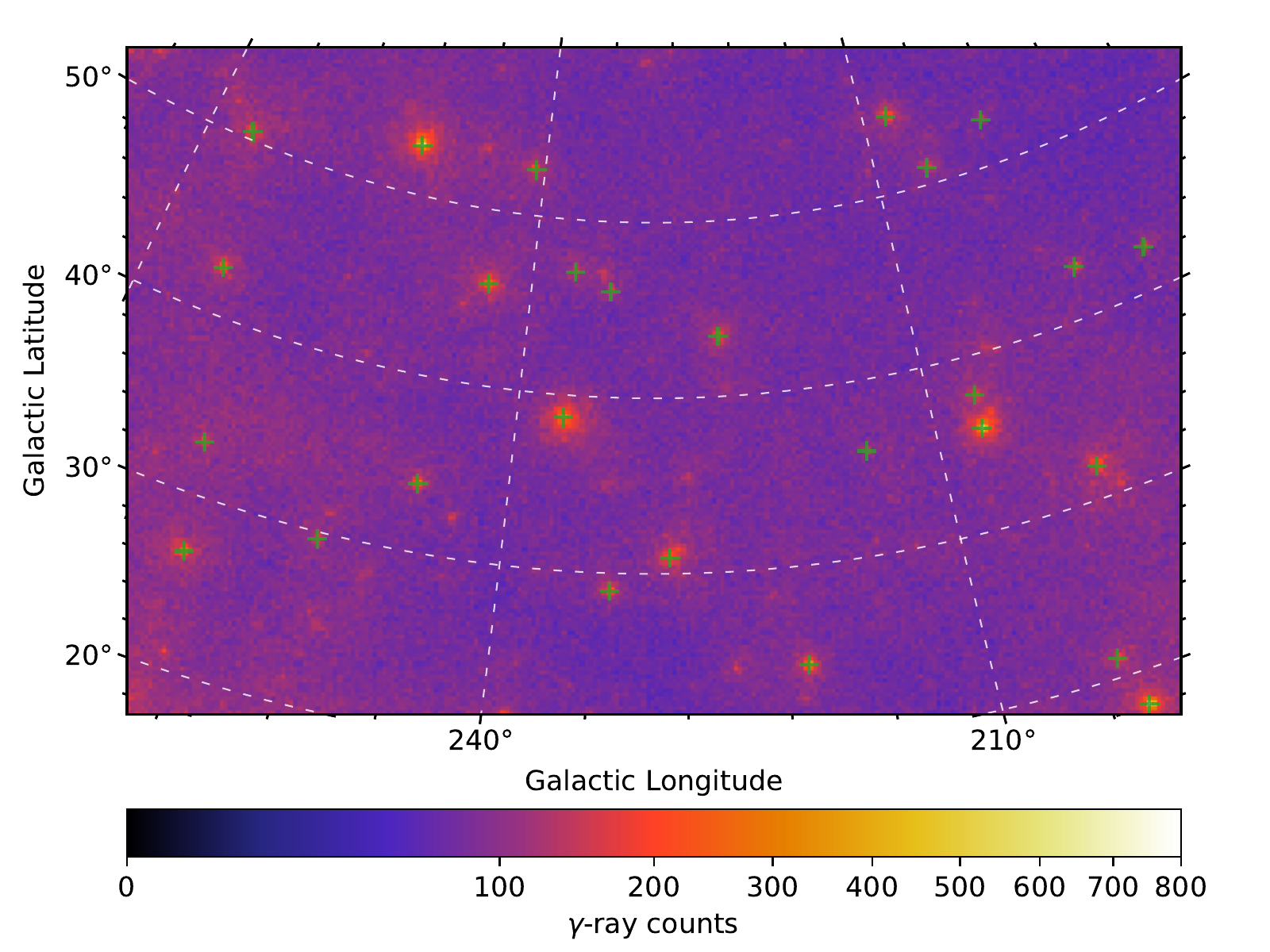}
    \includegraphics[width=0.48\textwidth]{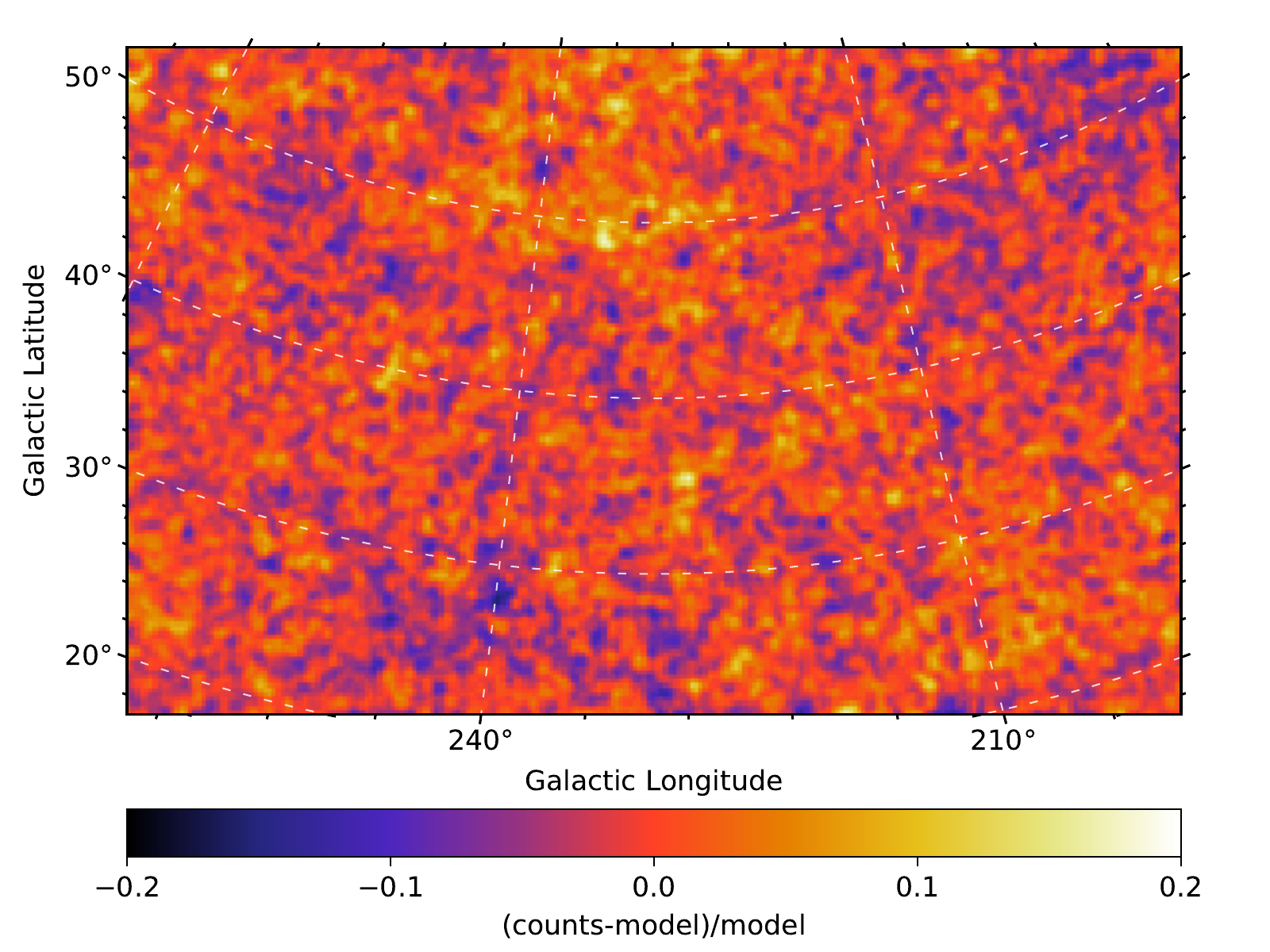}
    \caption{\label{fig:cmap}
        Count map (left) for \grs from 75~MeV to 100~GeV and residual map (right) showing the difference between the observed and modeled counts divided by the expected count map.
        The bright point sources which are fitted separately are marked as green crosses in the count map.
        The residual map is smoothed with a $1\deg$ Gaussian kernel to reduce statistical fluctuation.
    }
\end{figure*}

We choose a rectangular area in the carr\'{e}e projection centering at $(l, b) = (230\deg, 41\deg)$ as our ROI.
The photons in the ROI are partitioned into $240 \times 152$ pixels with the bin size of 0.25\deg, as shown in the Fig.~\ref{fig:cmap}, and 25 logarithmically spaced energy bins to build a count cube.

Throughout this work, {\tt fermitools} v1.0.0,\footnote{\url{https://fermi.gsfc.nasa.gov/ssc/data/analysis/software/}} the latest toolkit for \lat data analysis, is used.

\subsection{Components of the Galactic diffuse emission}\label{sec:analysis:gr_component}

\begin{figure}
    \centering
	\includegraphics[width=0.48\textwidth]{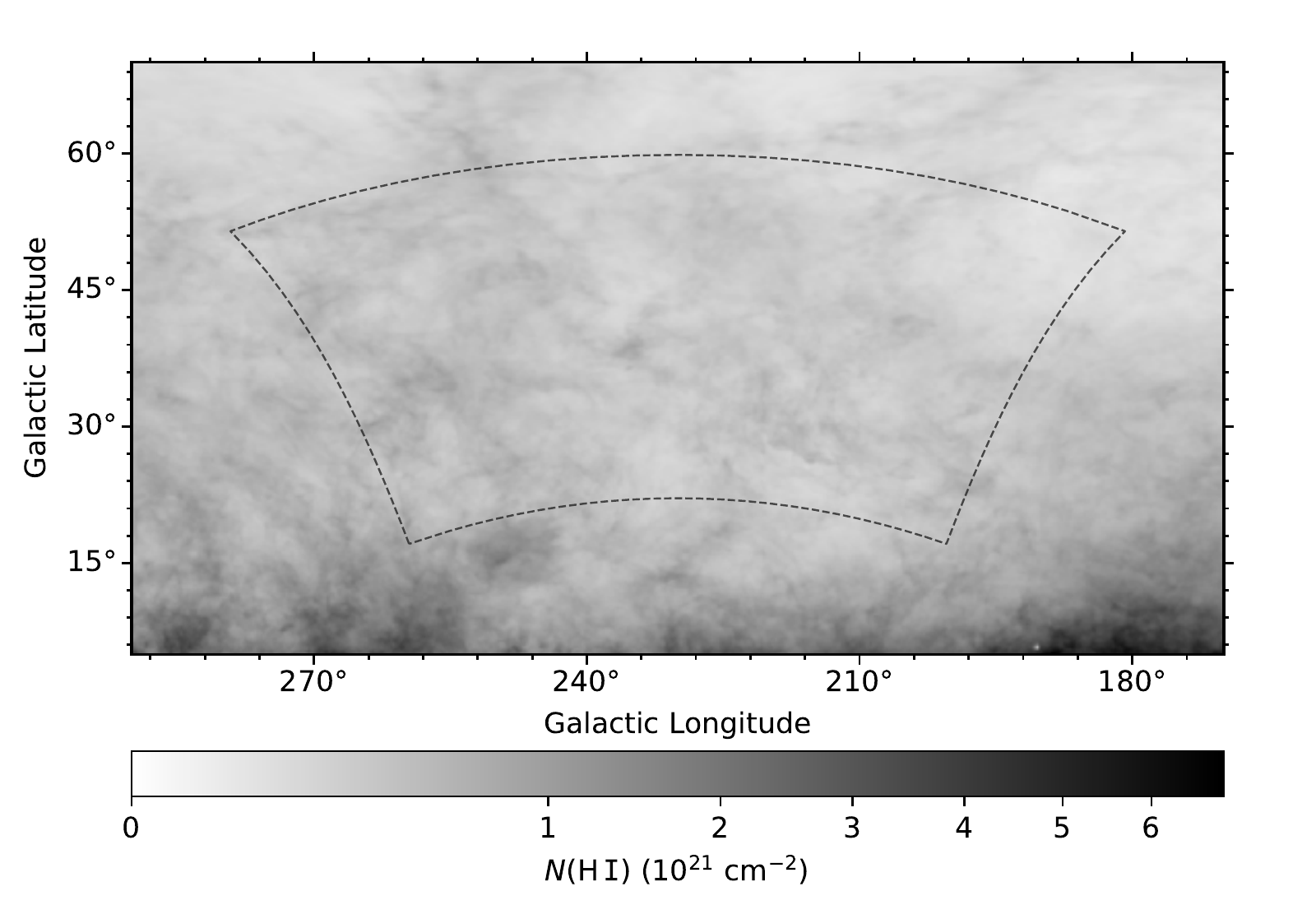}
    \caption{\label{fig:gas_Hi}
        The \hi column density in the source region.
        The spin temperature is assumed to be 125~K.
		The dashed line encloses the ROI of our analysis.
    }
\end{figure}

Galactic \gr diffuse emission originates from the interaction of CRs with interstellar gas and radiation field.
The decay of $\pi^0$ mesons and the electron bremsstrahlung are responsible for the former component, while the inverse Compton (IC) process contributes to the latter one.
Since all the diffuse emissions are merged into a single interstellar emission model {\tt gll\_iem\_v06.fits}~\citep{FermiGDE2016}, we need to replace it with its composition to derive the \gr spectrum associated with \hi.
The main procedure of making each component is very similar to~\cite{Shen2018_L5GEE} and will be described in the following.
To take into account the photons reconstructed inside the ROI but originated from sources outside, we define $170\deg \leq l \leq 290\deg$, $5\deg \leq b \leq 70\deg$ as our source region (SR), within which we make the templates.

The atomic hydrogen contributes to the majority of the gas in the SR~\citep{Abdo2009_3VMFM}.
We use the 21-cm hyperfine structure line data provided by the \hi $4\pi$ survey (HI4PI)~\citep{HI4PI2016}, as it provides a better angular resolution compared to its predecessor~\citep{LAB2005}.
Even though HI4PI covers a wide local standard of rest (LSR) velocity range from $-600~\kms$ to $600~\kms$, we exclude the data with $|v_{\rm LSR}| \geq 70~\kms$ following~\cite{PlanckOpacity2013} to eliminate the \hi emission from high-velocity clouds (HVCs) and extra-Galactic objects within our SR~\citep{Westmeier2018_MCWYC}, concerning no dust thermal emission~\citep{Wakker1997_7V8FM} or \gr emission~\cite{Tibaldo2015_BGTRD} of HVC has been found.
To calculate the \hi column density $N(\hi)$, we assume the spin temperature $T_{\rm S}=125~\rm K$ in our baseline model~\cite{Abdo2009_3VMFM}, and will try other $T_{\rm S}$ during the evaluation of systematic uncertainties.
The final $N(\hi)$ map is shown in Fig.~\ref{fig:gas_Hi}.

Although our ROI is chosen to exclude bright molecular clouds, there are still some CO emissions at the edge of the SR.
These clouds might influence our results in the lowest energy range due to the poor angular resolution.
The CO lines observed by the CfA telescope~\citep{Dame2001_F22PM} and optimized with moment masking method~\citep{Dame2011_YC149} are used in this work.
We integrate the CO brightness temperature in the SR to  construct the $W_{\rm CO}$ map.
The pixels sampled at 0.25\deg are linearly interpolated to 0.125\deg~\citep{Abdo2010_ALWUR}.
We notice that CfA survey only observes the CO emission at $|b| < 32\deg$, which does not cover all of the SR, however the completeness is proven with the emission from dust and \hi~\citep{Dame2001_F22PM}.
We also do not find significant CO clouds appear in the \planck TYPE 1 CO map~\cite{PlanckCO2014} but are unobserved by CfA survey in our SR.\footnote{There are indeed some point-like structures in this map with $W_{\rm CO} \lesssim 1~\rm K\;\kms$, however they seems related to extragalactic sources.}

Ionized gas is also a component of the interstellar gas with a typical volume-averaged free electron density $n \approx 0.01-0.1~\rm cm^{-3}$.
Since the diffuse warm ionized gas is 8 times more extended in scale height than \hi~\citep{Gaensler2008_DHPN2}, it will contribute to the gas column density in the SR.
Based on the \planck emission measure map in~\cite{PlanckFreefree2015}, we adopt the method detailed in~\cite{Westerhout1958_0EFE3,Ackermann2012_BRHCL} and the effective electron density $n_{\rm eff}=2~\rm cm^{-3}$~\citep{Sodroski1997_MZVVG} to make \hii column density N({\hii}).

Other than the gas of different phases directly traced by multi-wavelength observations, a missing component still exists in the total gas column density derived from dust thermal emission and \grs~\citep{Grenier2005_636PV}.
This extra component, known as the dark neutral medium (DNM), consists of the optically thick \hi and the \htwo without CO emission~\citep{Grenier2005_636PV,PlanckOpacity2011,Murray2018_B2FN0}.
Despite little CO emission is observed in SR, it is still possible that DNM exists.
We choose the latest 353~GHz dust opacity map~\citep{PlanckOpacity2016} as our dust tracer template $D(l,b)$ and derive the DNM template using the iterative method described below.
We first make an initial DNM map with all pixels being zero.
Then a linear combination of gas and DNM templates is calculated as the expected total gas column density, i.e.
\begin{eqnarray}
	M(l,b) 
	&=& D_{\rm DNM,prev}
	 + y_{\hi} \left[ N({\hi})+X_{\rm CO} W_{\rm CO} \right] \nonumber\\
	&+& y_{\hii} N({\hii}) + y_{\rm iso},
	\label{eqn:dust_model}
\end{eqnarray}
where $D_{\rm DNM,prev}$ represents the DNM map derived from previous iteration and $y_{\rm iso}$ is introduced to account for the residual noise and the uncertainty of dust map in the zero level~\citep{PlanckOpacity2013}.
The expected total density is fitted against the 353~GHz opacity map which minimizes the difference govern by
\begin{equation}
    \chi_{\rm dust}^2 = \sum_{l,b} \frac{\left[ D(l,b)-M(l,b) \right] ^2}{\sigma^2(l,b)},
    \label{eqn:chi2_dust}
\end{equation}
where $\sigma(l,b)$ is defined to be proportional to $D(l,b)$~\cite{Ade2015_UUM64,Tibaldo2015_BGTRD,Remy2017_7VYY4}.
Considering no CO emission appears in the ROI, we simply add the \hi and CO together with a fixed CO-to-\htwo conversion factor $X_{\rm CO}=0.9\times10^{20}~\rm K\;\kms$~\citep{Casandjian2015_K3EGW} in eq.(\ref{eqn:dust_model}) to make the fitting easier to converge.
After the optimization, the excess with more than $3\sigma$ deviation from the core of the residual map distribution is extracted as the new DNM template for the next iteration.
The fitting and extraction procedure continue until the $\chi^2$ in eq.(\ref{eqn:chi2_dust}) stabilizes, and the $D_{\rm DNM}$ in the last iteration is our final template.

The final part in the Galactic diffuse model is the IC radiation.
We adopt the same IC model as the one in the standard \lat Galactic model~\citep{FermiGDE2016}, which is calculated with the CR propagation code {\tt Galprop}\footnote{\url{https://galprop.stanford.edu/}}~\citep{Galprop1998,Strong2000_V6PA1,Porter2008_NEDZ6} using the parameter set named as ${\rm ^SY ^Z6 ^R30 ^T150 ^C2}$~\citep{Ackermann2012_L9WYG}.
Different IC models will also be analyzed as we evaluate the systematic uncertainties.

Loop I is a circle-like structure with a diameter of $\sim 100\deg$.
It was discovered in a survey of radio continuum~\citep{Large1962_Feature} and is also visible in the \gr band~\citep{Grenier2005_Estimate,Casandjian2009_High}.
Although its \gr emission is contributed by the IC process as well, it is not contained in the {\tt Galprop} IC model.
We include the Loop I in our analysis since it locates on the edge of our ROI.
We adopt a geometrical model~\cite{Wolleben2007_Model} using the parameters from~\citep{Ackermann2014_Spectrum} as our Loop I template.

\subsection{\gr analysis procedure}\label{sec:analysis:procedure}
Instead of adopting the correlation-based method in~\cite{Abdo2009_3VMFM}, we follow the well developed analysis scheme assuming the gas is transparent to \grs~\citep{Lebrun1983_QKGG3,Strong1988_8K8VL,Digel1996_MN2EM,Grenier2005_636PV,Abdo2010_ALWUR,Casandjian2015_K3EGW}.
The \gr intensity $I_\gamma$ in the direction of $(l,b)$ at the energy $E$ is given by
\begin{eqnarray}
    I_{\gamma}(l,b,E)
        &=& q_{\rm \hi}(E) \, \left [ N({\hi})(l,b) + X_{\rm CO} \, W_{\rm CO}(l,b) \right ] \nonumber\\
        &+& q_{\rm \hii}(E) \, N({\hii})(l,b)
         +  q_{\rm DNM}(E) \, D_{\rm DNM}(l,b) \nonumber\\
        &+& x_{\rm IC}(E) \, I_{\rm IC}(l,b,E)
         +  x_{\rm LoopI}(E) \, I_{\rm LoopI}(l, b)\nonumber\\
        &+& x_{\rm iso}(E) \, I_{\rm iso}(E) \nonumber\\
        &+& x_{\rm ps}(E) \, \sum_{j=1}^{n_{\rm ps,nf}} S_j(E) \, \delta(l-l_j,b-b_j) \nonumber\\
        &+& \sum_{k=1}^{n_{\rm ps,f}} S_k(E;\theta_k) \, \delta(l-l_k,b-b_k),
    \label{eqn:gr_model}
\end{eqnarray}
where $q$ stands for the \gr emissivity of the corresponding gas, and scaling factor $x$ is intended to fine tune the spectrum given in the map cube model.
Since no CO emission in the ROI, we combine the $W_{\rm CO}$ with the \hi column density using a fixed factor $X_{\rm CO} = 0.9\times10^{20}~\rm K\;\kms$~\citep{Casandjian2015_K3EGW}.
$I_{\rm iso}$ is the intensity of the isotropic background tabulated in {\tt iso\_P8R3\_CLEAN\_V2.txt}.
The sources inside the SR listed in the \lat 8-year point source list\footnote{\url{https://fermi.gsfc.nasa.gov/ssc/data/access/lat/fl8y/}} (FL8Y) are included, with the bright ones shown in Fig.~\ref{fig:cmap}.
The spectrum for each source is $S(E;\theta)$ with the parameters being $\theta$.
To limit the number of free parameters, sources with statistical significance smaller than 25 are merged into a single template based on the parameters given in the catalog and the others are left as individual templates.
The number of point sources with spectral parameters freed is $n_{\rm ps,f}$ and the number of the remaining is $n_{\rm ps, nf}$.

The expected \gr intensity given above is convolved with the \lat instrumental response functions (IRFs) with the {\tt gtsrcmaps}, and the binned likelihood fitting is performed using the {\tt pyLikelihood}~\citep{Mattox1996_DNQ4K,MINUIT1975}.
Since the uncertainty of the energy measurement will distort spectral parameters especially in the lower energy range, we take the energy dispersion correction into account for all the \gr emitting components except the isotropic background.\footnote{\url{https://fermi.gsfc.nasa.gov/ssc/data/analysis/documentation/Pass8_edisp_usage.html}}

We first perform a global fit before the bin-by-bin analysis, which helps to alleviate the overfitting problem and make the energy dispersion correction more accurate.
In the global fit, we choose the LogParabola spectral type for all the emissivities of gas, and optimize both the normalizations and the spectral indexes.
Because the spectral shape and spatial map of the IC emission are related to the CR electron distribution in the Milky Way, we only fit its normalization.
The intensity of the standard isotropic background is derived based on the standard Galactic interstellar model~\citep{FermiGDE2016}, so instead of just varying its prefactor in the fitting, we adopt a PowerLaw scaling factor to adjust the spectral index as well.
We set free normalizations of FL8Y sources with significance larger than 25.
The spectral shapes of the sources $\geq 35\sigma$ are also fitted.
Concerning the sources merged into a single template, we adopt a PowerLaw scaling factor to tune their prefactors and indexes as a whole.

\begin{figure}
    \centering
	\includegraphics[width=0.48\textwidth]{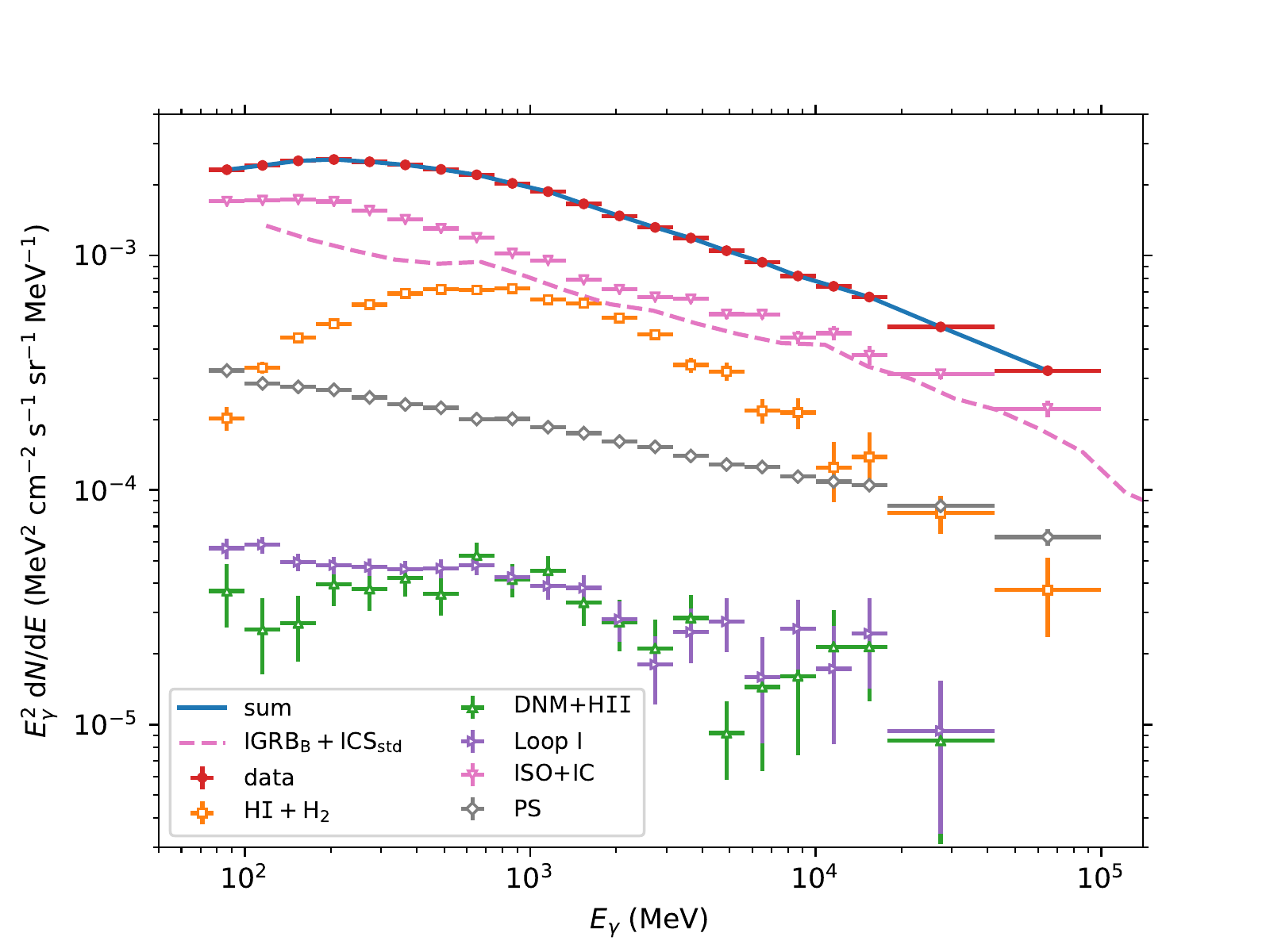}
    \caption{
        The intensity spectra for the observed counts and the components for our ROI in the baseline model.
        The uncertainties of the counts are only statistical, while the error bars of each component are obtained from the fittings.
        The pink dashed line shows the isotropic diffuse \gr background (IGRB) model B in~\cite{Ackermann2015_0PQU0} plus the IC spectrum from ${\rm ^SY ^Z6 ^R30 ^T150 ^C2}$ model.
    }\label{fig:spec_components}
\end{figure}

A bin-by-bin fitting is performed based on the resultant model in the global fit.
Since the inference in high energy range suffers from low statistics, we treat the six highest energy bins as two bins and fit three of them each time.
All the spectral indexes are kept fixed during the optimization.
Furthermore, we replace the indexes of DNM and \hii with that of \hi, because the first two are much weaker than \hi and their emissivities should have the same shape as \hi.
The normalization of IC model is also frozen to reduce the correlation with the isotropic background.

\begin{figure}
    \centering
	\includegraphics[width=0.42\textwidth]{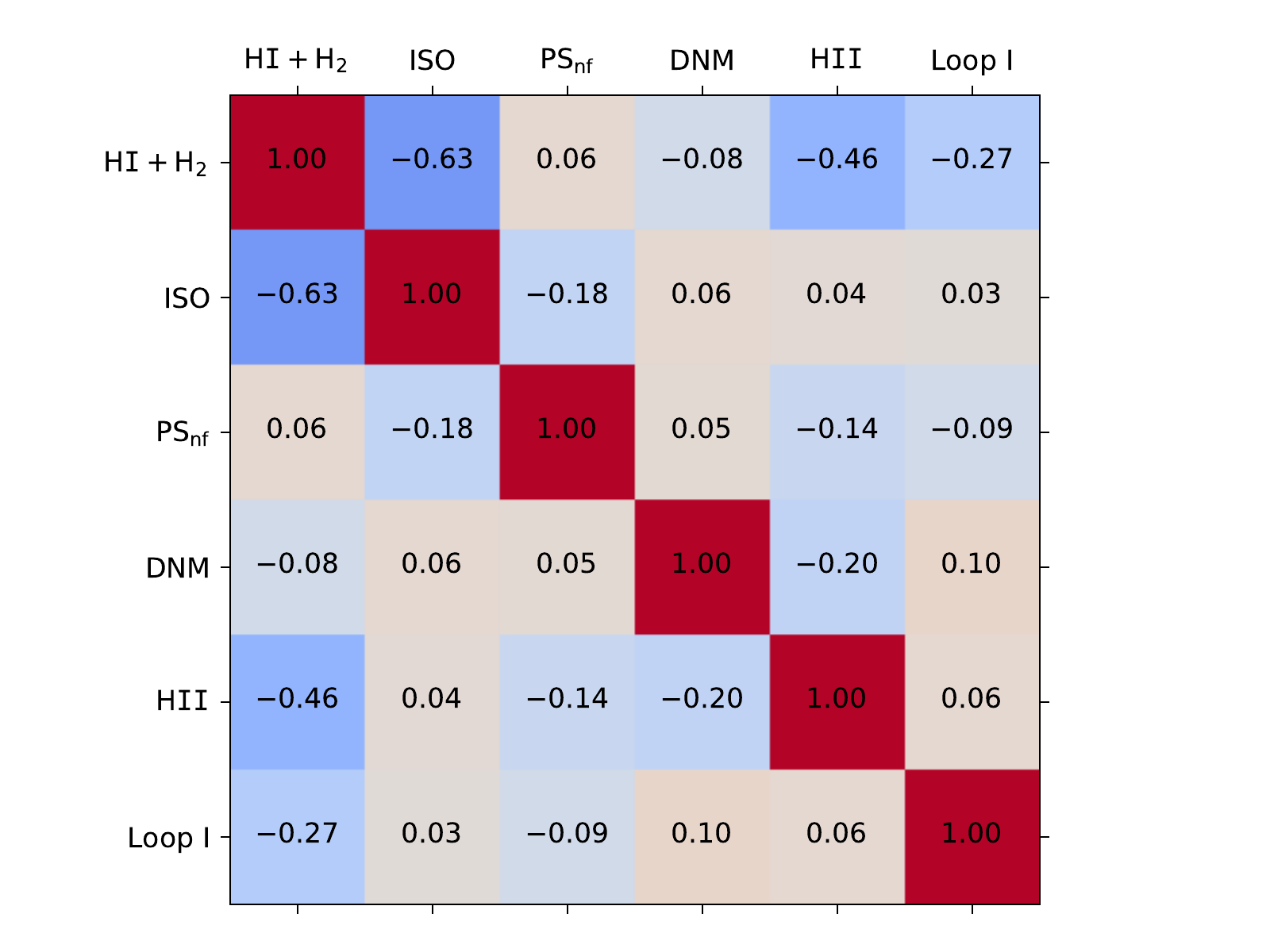}
    \caption{
        The correlation matrix between the normalizations of different large scale components in the baseline model derived in the global fitting.
        The $\rm PS_{nf}$ represents the single template for the weak point sources.
        We do not present the correlations of strong point sources, since they are fitted as individual point sources and are seldom correlated with the diffuse components.
    }\label{fig:corr_components}
\end{figure}

Before presenting the results, we will first do some fit quality checks.
Based on the best-fit parameters in each energy bin, we make a residual map (right panel of Fig.~\ref{fig:cmap}) and average intensities of different \gr emitting components (Fig.~\ref{fig:spec_components}).
To obtain the residual map, we subtract the sum of best-fit models in each energy bin from the observed count map, and then divide it by the predicted map.
The maps are smoothed with $1\deg$ Gaussian kernel to reduce the statistical fluctuation.
We do not find any significant structure in the residual map, with the minimum and maximum deviation being $-0.15$ and $0.17$ respectively.
In the intensity map, we adopted the average intensity of each component in the ROI along with its uncertainty obtained from the fittings.
We combine the spectra of isotropic and IC components since both of them are structureless in the ROI.
We also add the intensities of DNM and \hii together, considering that they are not as significant as other components and should have a similar spectral shape.
The uncertainty of the combined components is calculated by summing quadratically the errors of individual contributions.
The spectrum of observed counts and its statistical uncertainty in the figure are also given.
As shown in the figure, the model can well describe the observed count spectrum.
Since the \hi gas is anti-correlated with some of the diffuse components, isotropic background in particular, as depicted in Fig.~\ref{fig:corr_components}, the uncertainty of \hi gas spectrum is larger than Poisson ones.
Our $\rm ISO+IC$ intensity is larger than the isotropic diffuse \gr background (IGRB) model B\footnote{Model B is the largest IGRB model presented in~\citep{Ackermann2015_0PQU0}.} in~\cite{Ackermann2015_0PQU0} plus the IC spectrum from ${\rm ^SY ^Z6 ^R30 ^T150 ^C2}$ model (pink dashed line) by around $\sim 10\%-50\%$.
It might be explained by residual CR background\footnote{There is at most 50\% difference between the IGRB model B and the isotropic background {\tt iso\_P8R3\_CLEAN\_V2.txt}.} and and the different IC models adopted in this work from~\cite{Ackermann2015_0PQU0}.

\section{Results and Discussion}\label{sec:result_discussion}
\subsection{Results of the baseline model and the systematic uncertainties}\label{sec:result_discussion:result}

\begin{figure}
    \centering
	\includegraphics[width=0.48\textwidth]{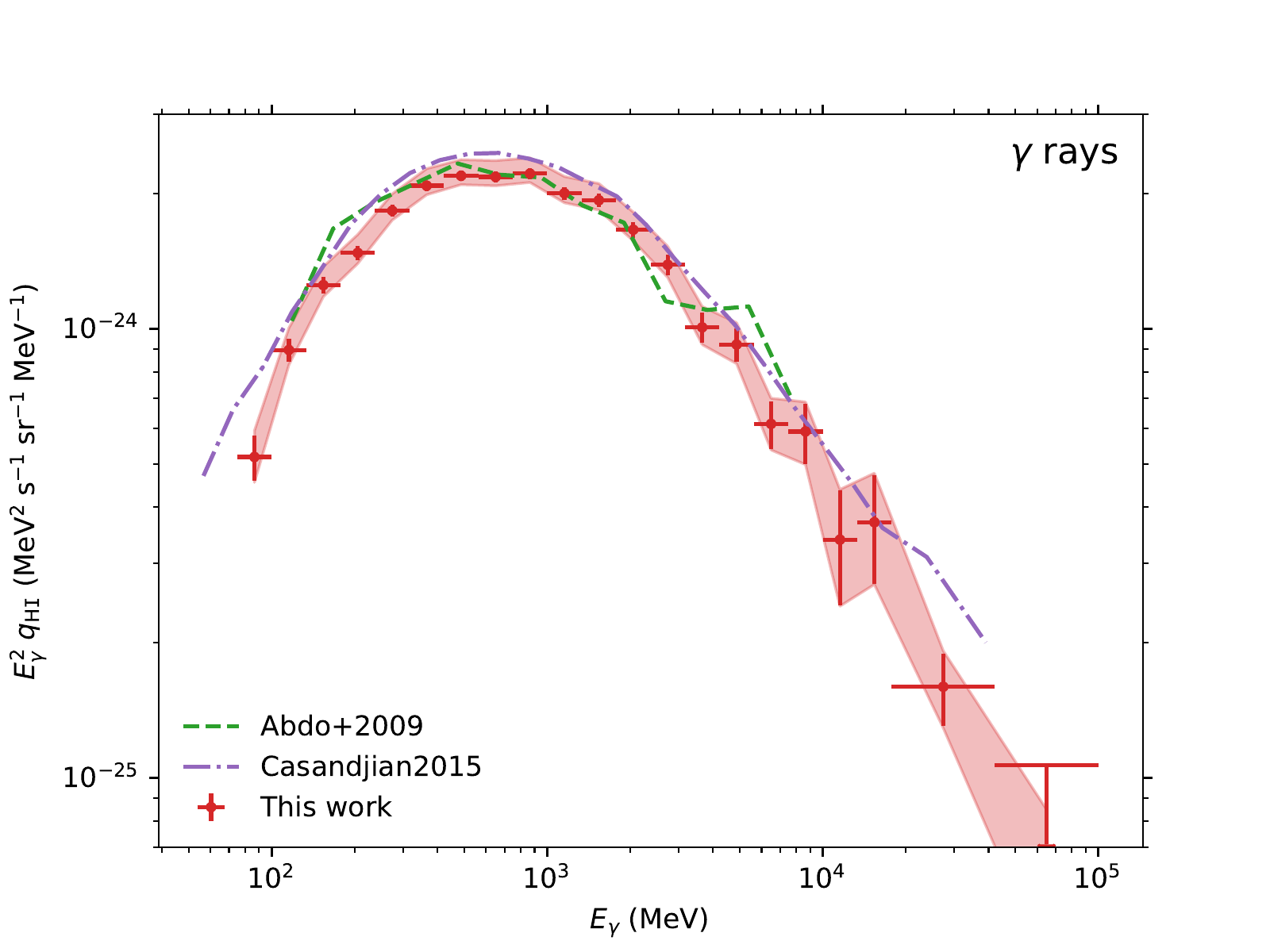}
    \caption{\label{fig:emissivity}
        \gr emissivity per \hi atom in the baseline model and its uncertainty.
        The red points and error bars indicate the best-fit values and $1\sigma$ statistical uncertainties in the baseline model.
        95\% upper limits are given when the TS value in that energy bin is smaller than 10.
        The red band shows the total errors including both the statistical and systematic uncertainties.
        The green dashed line and purple dot-dashed line illustrate the emissivity in~\citep{Abdo2009_3VMFM} and~\citep{Casandjian2015_K3EGW}, respectively.
    }
\end{figure}

The \gr emissivity per \hi atom in each energy bin is obtained using the best-fit spectral parameters, which is illustrated in Fig.~\ref{fig:emissivity}.\
The integral emissivities above 75~MeV and 100~MeV are $(1.63\pm0.08)\times10^{-26}~\rm ph\;s^{-1}\;sr^{-1}$ and $(1.46\pm0.06)\times10^{-26}~\rm ph\;s^{-1}\;sr^{-1}$ respectively.
Comparing with~\cite{Abdo2009_3VMFM} (green dashed line), which adopted a similar ROI to ours, spectral shape is similar but the integral is smaller, which might be caused by the updated background and templates.
Our emissivity is also consistent with~\cite{Casandjian2015_K3EGW} (purple dot-dashed line), which uses a larger ROI but older \gr data and gas tracers.

The \gr emissivity above is based on the templates described in Sec.~\ref{sec:analysis:gr_component} and the standard \lat IRFs.
In order to investigate the systematic uncertainties associated with them, we substitute the \gr emitting templates and also propagate the uncertainty on effective area in the following.
During the evaluation, the data analysis procedure is the same as that given in Sec.~\ref{sec:analysis:procedure}.

A uniform spin temperature $T_{\rm S} = 125~\rm K$ is used in the baseline model to convert the brightness temperature into the \hi column density.
A higher $T_{\rm S}$ means more electrons in hydrogen atoms are in the higher energy spin state, thus less absorption is experienced and smaller column density is expected.
We try three different $T_{\rm S}$ values 100~K, 200~K and $\infty$~K, which will decrease the \hi column density by $-1.5\%$, 2.1\% and 5.3\% on average respectively.
This type of uncertainty causes the emissivity to shift between $\sim -2\%$ and $\sim 8\%$.

The IC model is calculated based on a specific propagation parameters with {\tt Galprop}~\citep{FermiGDE2016}.
Different parameters will lead to different spectral and spatial shapes, and thus affect the emissivity.
We vary the IC model by using different {\tt Galprop} parameter sets~\citep{Ackermann2012_L9WYG,dePalma2013_T1HUP}, whose identifications are ${\rm ^SL ^Z4 ^R20 ^T150 ^C2}$, ${\rm ^SL ^Z10 ^R30 ^T\infty ^C5}$, ${\rm ^SS ^Z4 ^R20 ^T150 ^C2}$, and ${\rm ^SS ^Z10 ^R30 ^T\infty ^C5}$.
The IC templates only affect the emissivity in the high energy range, which leads to at most 2\% difference above $\sim 50~\rm GeV$.

The uncertainty of the effective area ($A_{\rm eff}$) dominates the instrument-related systematic uncertainties.
In our case, the largest relative uncertainty is 10\% at 31.6~MeV, decreases to 3\% at 100~MeV, stays at 3\% until 100~GeV, and then increases to 15\% at 1~TeV.
We use the bracketing $A_{\rm eff}$ method to propagate the uncertainty to the spectral parameters.\footnote{\url{https://fermi.gsfc.nasa.gov/ssc/data/analysis/scitools/Aeff_Systematics.html}}
To investigate the largest influence, we replace the $A_{\rm eff}$ with the upper and lower bound of the uncertainty for the sources with spectral indexes freed except the isotropic background and the merged template for weak sources.
It results in a $3\%-5\%$ change of the emissivity.

The total uncertainty including the statistical and systematic uncertainty is calculated with their root sum square,\footnote{We use $0.5(q_{\rm UL}-q_{\rm best})$ as the statistical error when TS value of that bin is below 10.} and is shown as a red band in Fig.~\ref{fig:emissivity}.

\subsection{LIS CR spectrum}\label{sec:result_discussion:cr_spec}

\begin{figure*}
    \centering
    \includegraphics[width=0.48\textwidth]{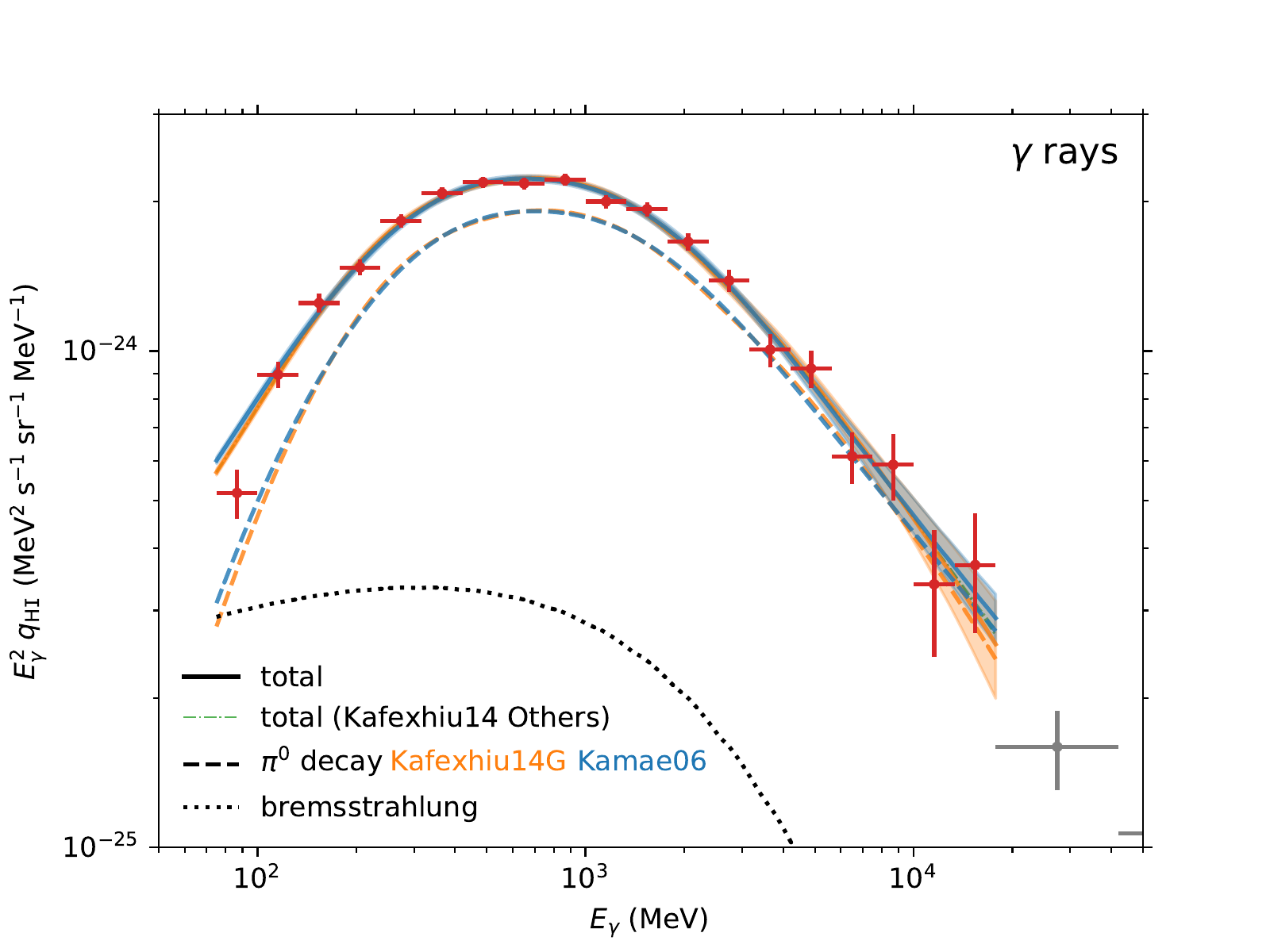}
    \includegraphics[width=0.48\textwidth]{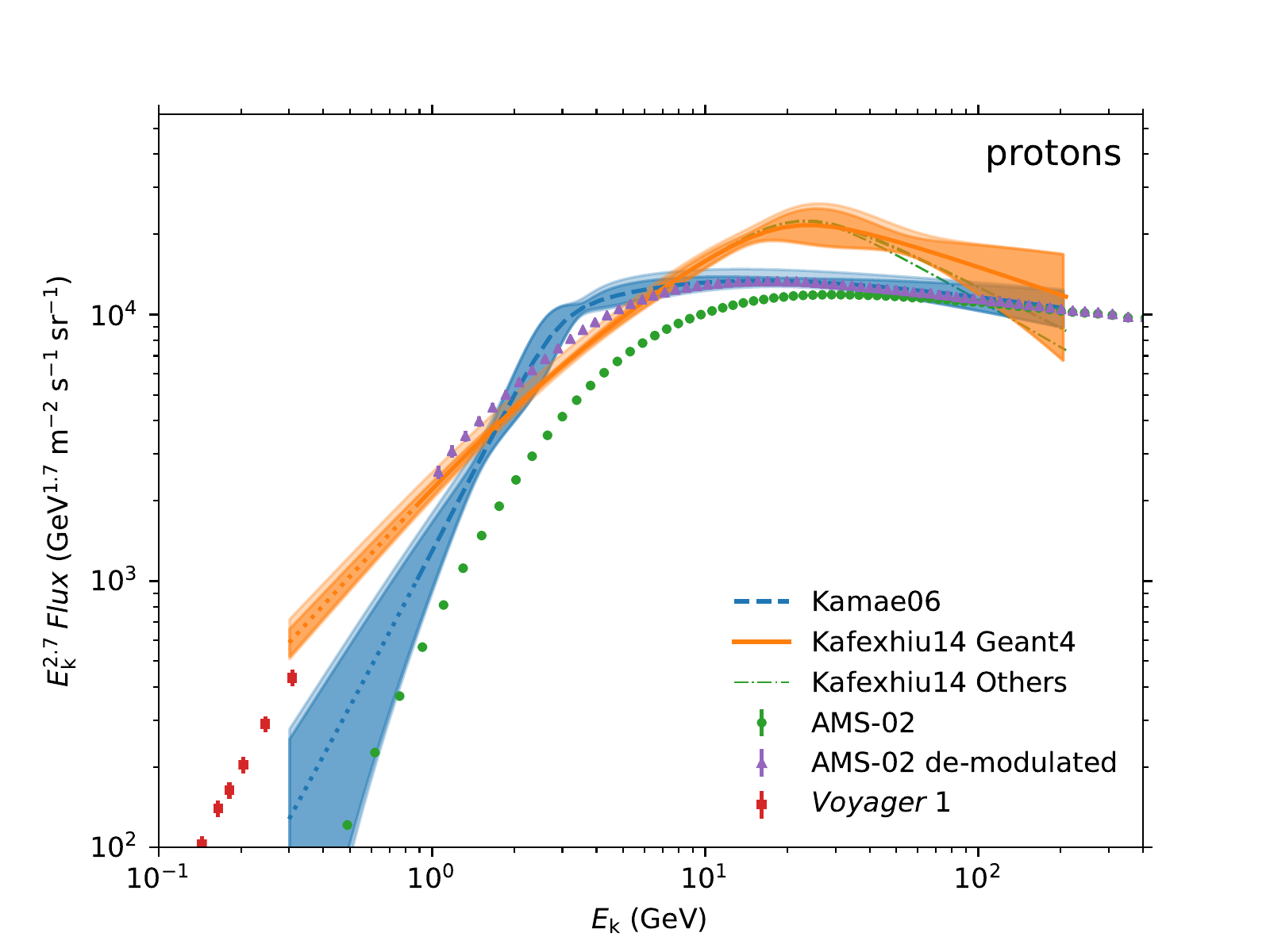}
    \caption{\label{fig:emiss_fit1}
        The \gr emissivity of \hi in the baseline model along with the best-fit \gr (left) and proton spectra (right).
        The data used in the fittings are drawn as red points and the others as gray ones.
        The results based on the cross section model from \cite{Kafexhiu2014_R6K7Z} and \cite{Kamae2006_UZAJA} are shown in orange and blue respectively.
        Only the statistical uncertainty of the baseline model is shown in left-hand figure.
        While in the right panel, the statistical and total errors are drawn with the dark and light color bands respectively.
        The extrapolated proton spectra are plot in dotted lines.
        The measurement from \emph{Voyager 1}~\citep{Cummings2016_W98KL} and AMS-02~\citep{Aguilar2015_DRNZF} are shown in red squares and green dots, while the de-modulated AMS-02 flux is shown in purple triangles.
        We also show the results using the parameterizations other than the {\tt EXPERIMENT+GEANT4} one with green dot-dashed lines in both figures.
    }
\end{figure*}

Since the \gr emissivity of interstellar gas comes from the $\pi^0$ decay and bremsstrahlung, a model consisting of the two emission processes is needed to fit the \gr observation and derive the CR spectrum.

The \gr production cross section in the $pp$ collisions is updated in~\cite{Kafexhiu2014_R6K7Z}.
It takes advantage of the published experimental data for the proton kinetic energy below 2~GeV and some sophisticated Monte Carlo codes in the higher energy.
We adopt the cross section parameterization {\tt EXPERIMENT+GEANT4} to account for the \gr emission from the process.
Since the interaction between a proton and a heavier nucleus may also produce \gr photons, we scale the cross section with an energy-dependent enhancement factor as in~\cite{Kafexhiu2014_R6K7Z}.
Because of the large systematic uncertainty of the cross section model, we also employ the widely-used cross section from~\cite{Kamae2006_UZAJA} and an enhancement factor of 1.78~\citep{Casandjian2015_K3EGW} as an alternative.
To avoid being cumbersome in the following, we define the \gr model containing the former cross section as KA14 model and containing the latter one as KK06 model.
The CR protons are assumed to follow a smoothly broken power law spectral shape~\citep{Strong2015_NELQ4}, i.e.
\begin{equation}
    {\rm d}F/{\rm d}p = A\,(p/p_0)^{-\alpha_1}\,[1+(p/p_{\rm br})^{(\alpha_2-\alpha_1)/\beta}]^{-\beta},
    \label{eqn:proton_spectrum}
\end{equation}
where $p$ is the momentum of a proton and $p_0$ is fixed to $3~{\rm GeV}\;c^{-1}$.
The normalization $A$, spectral indexes $\alpha_1$, $\alpha_2$ and break momentum $p_{\rm br}$ will be optimized and the smoothness factor $\beta$ will be fixed to 0.2.\footnote{If this factor is fitted, the improvement of $\chi^2$ is less than 0.4 and the derived parameters only change within 1$\sigma$ uncertainty.}

As to the bremsstrahlung emission, the cross section from~\cite{Strong2000_V6PA1} is employed.
We also include the bremsstrahlung emission from the CR electrons and positrons scattered by the heliums, which is a factor of 0.096 the abundance of hydrogen in the local ISM~\citep{Meyer1985_4Z670}.
Since the bremsstrahlung is the subdominant component in our energy range, we simply use the all-electron spectrum for PDDE model in~\cite{Orlando2018_2VP0Q}, which is well fitted to the directly measured electron spectrum and some synchrotron observations between 40~MHz and 20~GHz.

We fit CR proton spectrum using the \gr emissivity of baseline model below 17.8~GeV.
The data in the higher energy range are excluded due to the low statistics.
The best-fit \gr models and the resultant proton spectra based on the two cross sections are shown in Fig.~\ref{fig:emiss_fit1}.
Because the \gr data are only from 75~MeV, proton spectrum below $\sim 900~\rm MeV$ is not directly constrained by \gr emissivity~\citep{Kafexhiu2014_R6K7Z}.
An extrapolation down to the kinetic energy of 300~MeV is performed based on the best-fit models and is indicated with dotted lines in the right panel.
The statistical uncertainty of the proton spectra is shown in dark shaded regions, and the total errors including the systematic uncertainty propagated from the \gr emissivity is given with the light color band.
To compare with the direct CR observations, we plot the proton measurements from AMS-02~\citep{Aguilar2015_DRNZF} and \emph{Voyager 1}~\citep{Cummings2016_W98KL} with green dots and red squares respectively.
Also drawn in purple triangles is the de-modulated AMS-02 proton flux.
To derive the solar modulation potential, the non-parametric method in~\cite{Zhu2018_E303V} is adopted.
We assume a spline interpolation of LIS proton spectrum and fit the proton spectra with and without correction to the AMS-02 and \emph{Voyager 1} measurements.
It results in a potential of $\phi=0.57\pm0.04~\rm GV$.

The KK06 model gives a reasonable fit to the baseline emissivity, with the $\chi^2/$dof being 18.3/15.
The best-fit parameters of the proton spectrum are $A=(6.9\pm2.0)\times10^2\ {\rm m^{-2}\;s^{-1}\;sr^{-1}\;(GeV}\;c^{-1})^{-1}$, $\alpha_1=0.9\pm1.0$, $\alpha_2=2.85\pm0.07$, and $p_{\rm br}=4\pm1~{\rm GeV}\;c^{-1}$.
We find the spectral index after break matches that of AMS-02 between 45~GV and 336~GV, which is $2.849\pm0.002$.
This model provides a consistent proton spectrum with that observed by AMS-02 in the energy range where the solar modulation does not have strong impact.
The maximum deviation is $\lesssim 20\%$ above 15~GeV.
When we compare the result with the de-modulated proton spectrum, the difference drops to $2\%$ above 10~GeV which can be explained by the statistical and systematic uncertainties. 
A break at $\sim 4~\rm GeV$ in the best-fit model is also visible in the de-modulated spectrum.
At the energy of $\sim 300$~MeV, the \emph{Voyager 1} measurement is approximately 3 times the value of the extrapolated one, corresponding to a $\sim 2.5\sigma$ deviation considering the uncertainties of the spectral parameters.
This difference can be statistical or caused by the uncertainty of the cross section.
If the first case is true, it suggests no strong modulation in the local ISM~\citep{Stone2013_G966N}.

The KA14 model can explain the \gr observation as well, whose $\chi^2/$dof is 16.6/15 and the parameters are $A=(5.5\pm0.1)\times10^2\ {\rm m^{-2}\;s^{-1}\;sr^{-1}\;(GeV}\;c^{-1})^{-1}$, $\alpha_1=2.20\pm0.08$, $\alpha_2=3.1\pm0.3$, and $p_{\rm br}=20\pm11~{\rm GeV}\;c^{-1}$.
The best-fit spectrum has different shape from the direct measurement above 10~GeV, which is also found in~\cite{Neronov2017_5UWT9}.
But concerning the large statistical errors in the break energy and the high-energy break, a spectral shape may still be consistent with the AMS-02 observation.
The predicted proton flux is approximately $1.4-1.8$ times the data of AMS-02 at $2-100~\rm GeV$ with the maximum deviation shown at the break energy.
The difference decreases to 18\% at $\sim 180~\rm GeV$.
We also try the other $pp$ collision cross-section parameterizations given in~\cite{Kafexhiu2014_R6K7Z}, which are mainly different from the {\tt EXPERIMENT+GEANT4} one when the kinetic energy of proton is larger than 50~GeV, and find that their predictions are even softer after break and still can not solve the current problem (shown in green dotted-dashed lines in Fig.~\ref{fig:emiss_fit1}).
When compared with the de-modulated spectrum below 10~GeV, the difference decreases to at most 20\%.
The extrapolation exceeds the measurement of \emph{Voyager 1} by $\sim 40\%$ at 300~MeV, which is $\sim 2\sigma$ larger.
If it is the case, either a mild bending below $\sim 900~\rm MeV$ is needed or the CR in the local ISM is slightly more than that observed by \emph{Voyager 1}.

\section{Summary}\label{sec:summary}
The \gr observation can be used to derive the \gr emissivity of interstellar gas and thereby the CR spectrum.
We choose a mid-latitude Galactic region as our ROI in this work to investigate the LIS CR spectrum.
Using the recent version of \lat data~\citep{Bruel2018_1LQ9V}, most complete point source catalog as well as the up-to-date multi-wavelength survey of interstellar gas~\citep{HI4PI2016,PlanckOpacity2016}, we obtained the \gr emissivity of \hi gas and its systematic uncertainties, which are illustrated in~Fig.~\ref{fig:emissivity}.
Then two \gr production cross sections of $pp$ interaction, the commonly used one from~\cite{Kamae2006_UZAJA} and the up-to-date one from~\cite{Kafexhiu2014_R6K7Z}, are adopted to convert the emissivity into the CR spectrum.

Even though the two models can both provide reasonable fits to the data, they yield different proton spectra.
The discrepancy between the spectra is $\lesssim 50\%$.
It suggests a significant influence of cross section on reconstructing the proton spectrum.

The KK06 model gives a spectrum rather consistent with the AMS-02 measurement but smaller than the \emph{Voyager 1} measurement.
The spectral index above the break is $2.85\pm0.07$, which is consistent with the result from AMS-02~\citep{Aguilar2015_DRNZF}.
There is $\lesssim 20\%$ deviation between the predicted spectrum and the AMS-02 measurement above 15~GeV, and the difference becomes as small as $2\%$ if we compare the prediction with the de-modulated data.
A break at $p=4\pm1~{\rm GeV}\;c^{-1}$ shown in our result is also visible in the de-modulated spectrum.
An index of $0.9\pm1.0$ is predicted in the low energy range.
If an extrapolation is performed down to $\sim 300~\rm MeV$, the proton flux is only about 33\% of the \emph{Voyager 1} measurement~\citep{Cummings2016_W98KL}, corresponding to a $\sim 2.5\sigma$ deviation.

The KA14 model yields a spectrum that deviates from the direct measurement in high energy (see also~\cite{Neronov2017_5UWT9}).
Specifically, about $1.4-1.8$ times the amount of directly measured protons are required between 2~GeV and 100~GeV.
The difference becomes $\lesssim 20\%$ below 10~GeV when it is compared with the de-modulated spectrum.
A break at $20\pm11~{\rm GeV}\;c^{-1}$ is needed, with the indexes before and after the break being $2.20\pm0.08$ and $3.1\pm0.3$, respectively.
The extrapolation exceeds the \emph{Voyager 1} measurement by $\sim 40\%$ at $\sim 300~\rm MeV$.

Nowadays, based on the \gr observation, a CR spectrum roughly resembling the direct measurement can be obtained, however the systematic uncertainty on the cross section (also shown in~\cite{Casandjian2015_K3EGW}) still prevents us from accurately determining the spectral parameters of CR protons.
The situation is expected to change once a more accurate cross section model is available.

%%%%%%%%%%%%%%%%%%%%%%%%%% ACKNOWLEDGEMENT %%%%%%%%%%%%%%%%%%%%%%%%%%%
\begin{acknowledgments}
    We acknowledge the help from C.-R.~Zhu and very helpful comments from the referees.
    We use the {\tt NumPy} \citep{numpy2011}, {\tt SciPy}\footnote{\url{http://www.scipy.org}}, {\tt Matplotlib} \citep{matplotlib2007}, {\tt Astropy} \citep{astropy2018} and {\tt iminuit}\footnote{\url{https://github.com/iminuit/iminuit}} packages and the SIMBAD \citep{SIMBAD2000} database during our data analysis.
    This work is supported by National Key Program for Research and Development (2016YFA0400200), the National Natural Science Foundation of China (Nos. 11433009, 11525313, 11722328), and the 100 Talents program of Chinese Academy of Sciences.
\end{acknowledgments}

%%%%%%%%%%%%%%%%%%%%%%%%%%% BIBLIOGRAPHY %%%%%%%%%%%%%%%%%%%%%%%%%%%%%
%\bibliography{mybibs} % bibtex file name (w/o suffix)

\end{document}